\documentclass[a4paper,10pt]{article}
\usepackage[english]{babel}
\usepackage{color}
\usepackage[margin=2cm,tmargin=2cm]{geometry}
\usepackage[nottoc]{tocbibind}
\usepackage{amsmath}
\usepackage{graphicx}
\usepackage{breqn}
\usepackage{color}
\usepackage{mathtools}
\usepackage{multicol}
\usepackage{url}
\usepackage{fancyhdr}
\usepackage{csquotes}
\usepackage{wrapfig}
\usepackage[dvipsnames]{xcolor}
\usepackage{lipsum}
\usepackage{atbegshi}
\graphicspath{{Images/}{../Images}}
\usepackage{subfiles}
\usepackage[toc,page]{appendix}
 \usepackage{subcaption}
\usepackage{blindtext}

\usepackage[acronym]{glossaries}
\makeglossaries
\usepackage{acronym}
\usepackage{titling}
\usepackage{lipsum}
\usepackage{siunitx}

\pretitle{\begin{flushleft}\LARGE\bfseries}
\posttitle{\par\end{flushleft}}
\preauthor{\begin{flushleft}\Large}
\postauthor{\end{flushleft}}
\predate{\begin{flushleft}}
\postdate{\end{flushleft}}

\title{Scintillating Fibre Detector for the Mu3e Experiment \large Contribution to the 25th International Workshop on Neutrinos from Accelerators}

\author{\normalsize\textbf{R.M. Amarinei\textsuperscript{1} for the Mu3e Collaboration}, \\ \vspace{0.3cm}
~\textsuperscript{1}Department of Nuclear and Particle Physics, University of Geneva, 
24 rue du Général-Dufour, Geneva, CH\\ 
~~~Email: robert.mihai.amarinei@cern.ch 
}

\begin{document}
\maketitle
\vspace{-0.7cm}
\hrule
\vspace{1cm}
\begin{abstract}
    We present a compact scintillating fibre timing detector developed for the Mu3e experiment. Mu3e is one of the flagship experiments of the Swiss particle physics scene, aiming to search for the charged lepton flavour violating “neutrinoless” muon decay $\mu^+ \rightarrow e^+e^-e^+$. Mu3e is planned to start taking data in 2025 at the Paul Scherrer Institute in Switzerland, using the most intense continuous surface muon beam in the world (10$^8$ muons per second).
    
    At the University of Geneva, together with partners from ETH Zurich, we are developing a scintillating fibre detector formed by staggering three layers of 250 $\mu m$ diameter round scintillating fibres. The fibre ribbons are coupled at both ends to multi-channel silicon photo-multiplier arrays, which are read out with the MuTRiG ASIC, specifically developed for this experiment.
    
    This presentation is focused on the performances of the scintillating fibre detector, notably on the time resolution around 250 ps, the efficiency greater than 97\% and the spatial resolution of $\sim$ 100 $\mu m$. In this presentation, we also include the challenges overcome to build this very thin scintillating fibre detector, having a thickness smaller than 0.2\% of the radiation length. Furthermore, we discuss the operation and performance of the MuTRiG ASIC, used for reading out the 3072 channels of the fibre detector.
    \end{abstract}
    
    \newpage

\section{Motivation for a Scintillating Fibre Detector}

The Mu3e experiment is designed to search for the charged lepton flavor violating decay of a muon into two positrons and an electron (\( \mu^+ \to e^+ e^- e^+ \)). This decay is strongly suppressed in the Standard Model (SM) with massive neutrinos, but is predicted by several extensions, making it a sensitive probe for physics beyond the SM. 

The Mu3e experiment will take place at the Paul Scherrer Institute (PSI), utilizing a high-intensity muon beam that delivers up to \( 10^8 \) muon stops per second. The high particle detection rates require overcoming substantial combinatorial background (for instance two Michel decays where one positron undergoes Bhabha scattering in the detector) and radiative background such as the radiative decay with internal conversion (\( \mu^+ \to e^+ e^- e^+ \nu_e \nu_\mu \)). To identify rare signal events among these backgrounds, Mu3e imposes stringent performance requirements on its detectors. It demands exceptional momentum resolution ($\leq 0.5$ \si{MeV/c}), time resolution ($\leq$ 500 \si{\pico\second}), and spatial resolution ($\sim$ 200 \si{\micro\meter}) to distinguish signal events with high precision. Furthermore, to minimize multiple scattering and enhance momentum resolution, the detectors must operate with an extremely low material budget.  

The primary technology for tracking in Mu3e is based on High-Voltage Monolithic Active Pixel Sensors (HV-MAPS), which provide thin, high-resolution pixel detection~\cite{HVMAPS}. The pixel sensors have a thickness of 50 or 70 \si{\micro\meter}, with a 23 \si{\micro\meter} spatial resolution and an efficiency of 99\%. While this technology performs very well for tracking and position determination, its time resolution is around 20 \si{\nano\second}. This time precision is not enough for the high particle rate of the Mu3e experiment, leading to the necessity of additional detectors specifically built for timing~\cite{Corrodi}.

While advancements in pixel technology have achieved impressive resolutions down to the picosecond range, scaling these technologies to the large areas required for Mu3e Phase I is not feasible within the available timeframe (first physics data in 2026). Therefore, the experiment will use more mature technologies based on scintillating materials: at the centre of the detector, there will be a very thin Scintillating Fibre (SciFi) detector, while on the outside, there is going to be a thicker (but faster) scintillating tile detector. This proceeding explores the development, integration, and expected performance of the SciFi detector.

\begin{figure}[htb!]
    \centering
    \includegraphics[width=0.7\linewidth]{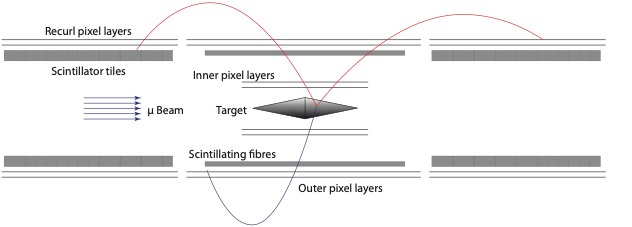}
    \caption{A detailed view of the Mu3e detector layers, with the muon beam directed toward the target positioned at the Mu3e core~\cite{Mu3e:Cite}.}
    \label{fig:Mu3eGeneralScheme}
\end{figure}

\section{The SciFi Detector}

The SciFi detector is located at the core of Mu3e (Fig.\ref{fig:Mu3eGeneralScheme}), positioned as the second innermost detector, just outside the vertex detector. It must be both fast and thin, and due to the limited physical space near the centre, the mechanical structure of the SciFi is designed with a highly constrained shape.

\begin{figure}
    \centering
    \includegraphics[width=0.44\linewidth]{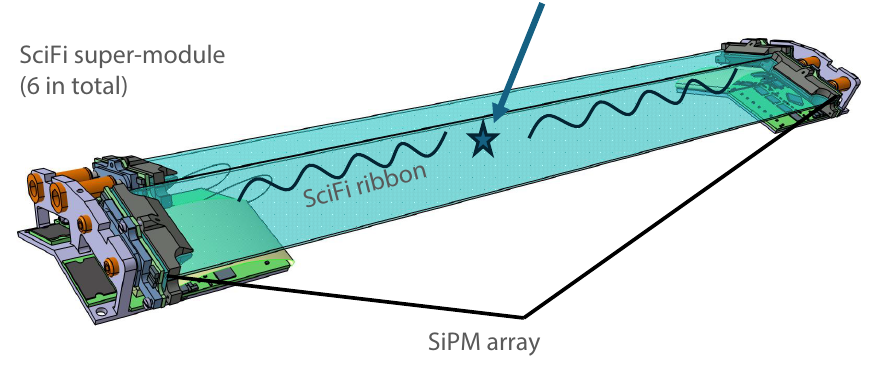}
    \includegraphics[width=0.55\linewidth]{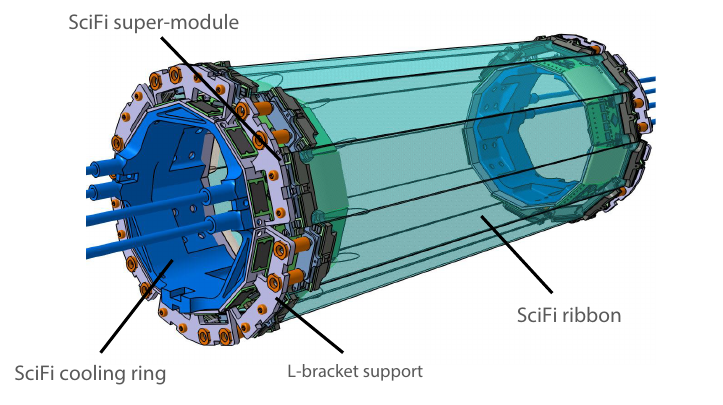}
    \caption{Left: A SciFi supermodule containing two ribbons with their readout boards and SiPM array mounted on an L-bracket. Right: CAD rendering of the full SciFi mounted on the cooling ring (deep blue). }
    \label{fig:CadSchemes}
\end{figure}

The building block of the SciFi detector is the so-called (scintillating fibre) ribbon. Each ribbon is constructed from 3 layers of 250 \si{\micro\meter} staggered scintillating fibres. A single ribbon has a total thickness of 720 \si{\micro\meter}, corresponding to a material budget of  $0.2\%$ of a radiation length.   The way in which particles are detected in a ribbon is depicted schematically in Fig.~\ref{fig:CadSchemes} — left. If a particle (blue arrow) interacts with the ribbon, the scintillating material is excited, releasing several photons upon de-excitation. These photons travel toward the ends of the ribbon, where they are detected by two 128 channel Silicon Photomultipliers (SiPMs). 
Below the SiPM arrays, there is the SciFi Module Board (SMB), the readout frontend used for data readout. The SMB is placed on an L-bracket which plays a very important role, both as a mechanical structure for the SciFi and as a cold element,  for the cooling of the SMBs and SiPMs.

The SciFi has a modular design composed of 6 independent supermodules, each of them containing two ribbons (Fig.~\ref{fig:CadSchemes} — right). 
The entire SciFi detector comprises therefore a total of 12 ribbons, arranged to provide full \( 4\pi \) coverage. The part holding together all the 12 ribbon modules is the SciFi cooling ring, which plays a central role in the cooling of the readout electronics and SiPMs.  The cooling ring has inner channels through which silicon oil at a temperature of -20$^\circ$C is circulated. This provides sufficient cooling for the readout boards and mitigates the radiation damage to the SiPMs placed in the highly radiative environment just outside the target, where $10^8$ muon stops happen per second.

\subsection{Performance of the SciFi}

The performance of the SciFi detector has been extensively studied, with several doctoral theses focusing on this technology~\cite{Corrodi,DemetsThesis,Damyanova}.
As discussed in detail in~\cite{Demets}, the performance of several ribbons was tested using surrogate electronics in a dedicated telescope setup, which provided a reference time. The results demonstrate that the SciFi ribbon, made of SCSF-78 plastic scintillator, achieves a time resolution of $\sim$ 250 \si{\pico\second}; twice as good as the initial requirements of Mu3e. For future upgrades, it has also been shown that a further improvement in time resolution 
 ($\sim$200 \si{\pico\second}) can be achieved using a 4-layer ribbon made of NOL-11 material. Another critical parameter studied during the beam tests is the efficiency of the module. This has been investigated in several configurations and is consistently higher than 97\%.

\begin{figure}
    \centering
    \includegraphics[width=0.45\linewidth]{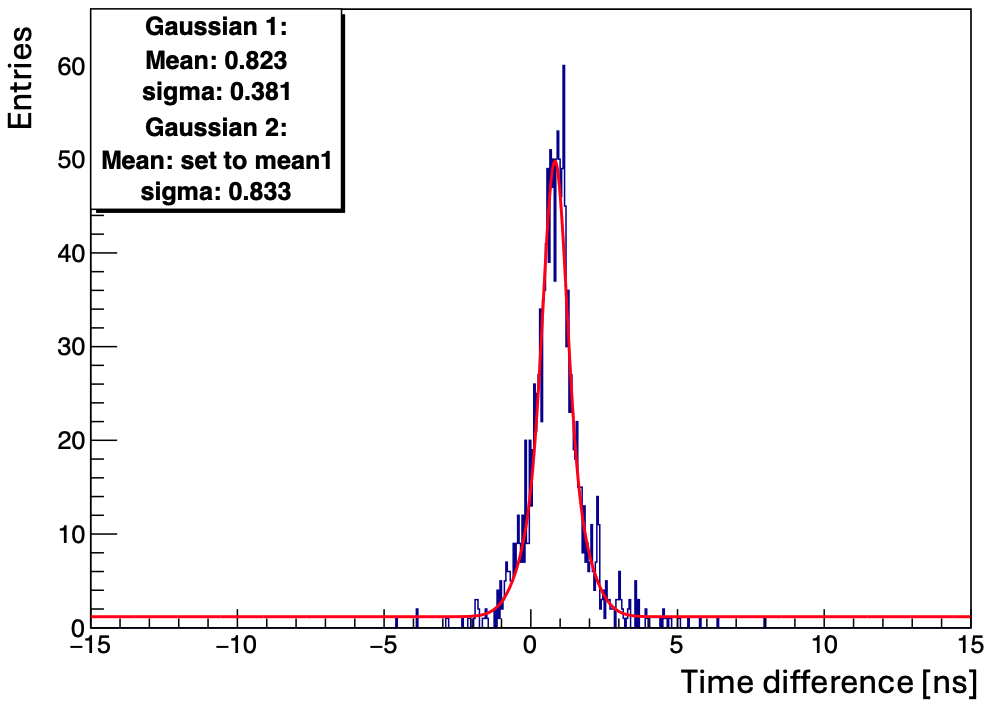}
    \includegraphics[width = 0.42\linewidth]{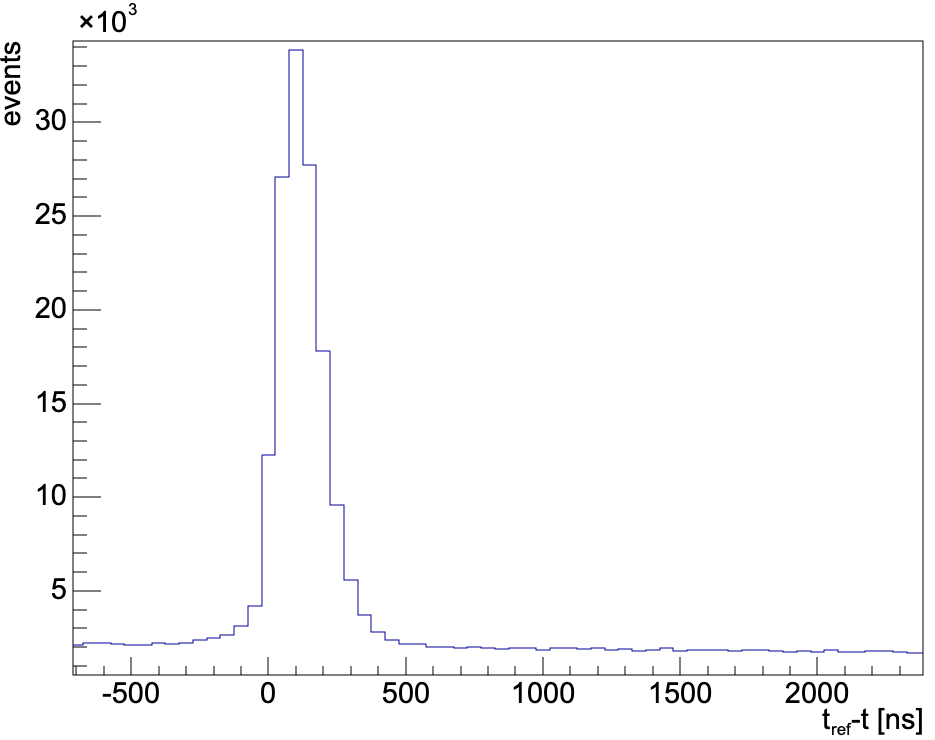}

    \caption{Left: Time difference between signal in coincidence between two ribbons. Right: Time correlation between one ribbon and two pixel modules.}
    \label{fig:MuTRiG_stud}
\end{figure}

Following these beam tests, studies have been conducted using the final readout electronics based on the MuTRiG ASIC~\cite{MuTRiG_cite}. These tests were performed in a further dedicated beam test setup together with the final pixel module to be employed in Mu3e. This marked the first time in which two final Mu3e technologies were tested together. 
Figure~\ref{fig:MuTRiG_stud} shows the results from a recent beam test, where pixel detectors (quad-modules) were tested alongside two final SciFi ribbons. The left panel of the figure displays the time difference between two consecutive SciFi ribbons, with the mean-time resolution of the two modules measured at $\sim$ 380 \si{\pico\second}. This apparent increase in time resolution is expected, as the uncertainty of the measurement is affected by the uncertainty of each individual SciFi ribbon.  Additionally, since the curve is closely centred around zero, it shows that the four boards (two per ribbon) are well synchronized, with a small shift attributed to jitter. 
The right panel of Fig.\ref{fig:MuTRiG_stud} illustrates the time difference between hits in coincidence for four pixel detectors and one SciFi ribbon. This plot conclusively demonstrates that the latest version of the DAQ and front-end electronics operates correctly across multiple Mu3e detectors.

\section{Status}

\begin{figure}[]
    \centering
    \includegraphics[width=0.6\linewidth]{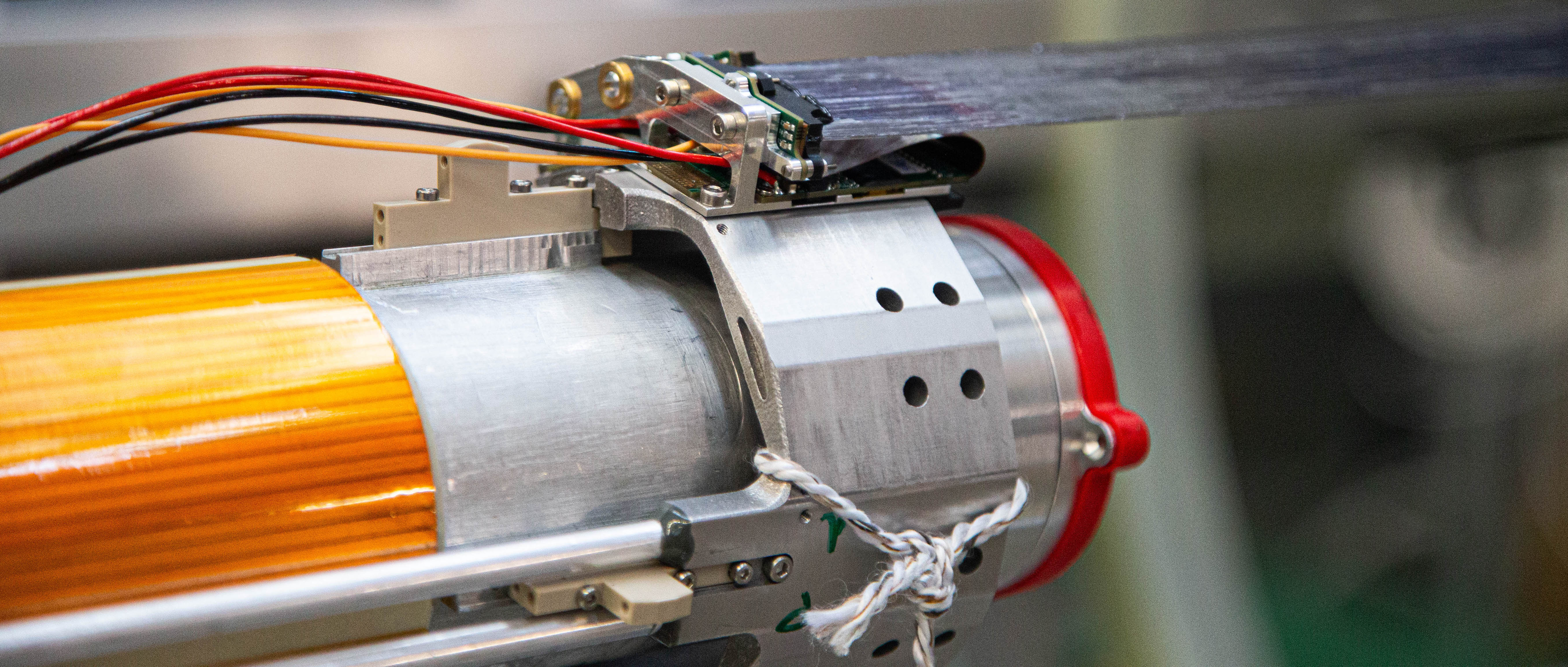}
    \caption{Close up of the SciFi cooling ring and first supermodule. Both of them are in their final version, ready to be used for the final Mu3e experiment. }
    \label{fig:PrettyPhoto}
\end{figure}

Currently, half of the SciFi modules are complete and ready for installation, while the remaining half is undergoing qualification. There are no remaining bottlenecks for the SciFi technology, and it is anticipated that the first set of cosmic ray data, taken in coincidence with the pixel detector, will be collected in the first quarter of 2025. Figure~\ref{fig:PrettyPhoto} provides a close-up view of one of the supermodules installed on the cooling ring.

At present, the SciFi cooling system has been successfully installed on the Mu3e cage. Performance tests of the cooling system have demonstrated its ability to maintain a stable temperature of -18$^\circ$C while dissipating the 60~W heat generated by the SciFi electronics. A full commissioning of the system is planned for the coming months, marking the final step before the SciFi modules are installed and operated within the Mu3e detector cage. For reference, the left side of Fig.~\ref{fig:PrettyPhoto} shows the copper bars mounted directly on the beam pipe, which provide low-voltage connections to the central detectors of Mu3e.

\bibliographystyle{unsrt}
\bibliography{references.bib}
\end{document}